\magnification 1200
\centerline {\bf  Can the Quantum Measurement Problem be resolved }
\vskip 0.3cm
\centerline {{\bf within the framework of Schroedinger Dynamics?}\footnote*{Lecture 
given at the J. T. Lewis Memorial Conference, held in Dublin, June 14-17, 2005}}
\vskip 1cm
\centerline {{\bf by Geoffrey Sewell}\footnote{**}{e-mail: g.l.sewell@qmul.ac.uk}}
\vskip 0.5cm
\centerline {\bf Department of Physics, Queen Mary, University of London}
\vskip 0.5cm
\centerline {\bf Mile End Road, London E1 4NS, UK}
\vskip 1cm
\centerline {\it To the Memory of John Lewis}
\vskip 1cm
\centerline {\bf Abstract}
\vskip 0.5cm
We formulate the dynamics of the generic quantum system $S_{c}$ comprising a 
microsystem $S$ and a macroscopic measuring instrument ${\cal I}$, whose pointer 
positions are represented by orthogonal subspaces of the Hilbert space of its pure states. 
These subspaces are the simultaneous eigenspaces of a set of coarse grained 
intercommuting macroscopic observables and, most crucially, their dimensionalities are 
astronomically large, increasing exponentially with the number, $N$, of particles 
comprising ${\cal I}$. We formulate conditions under which the conservative dynamics 
of  $S_{c}$ yields both a reduction of the wave packet describing the state of $S$ and a 
one-to-one correspondence, following a measurement, between the pointer position of 
${\cal I}$ and the resultant eigenstate of $S$; and we show that these conditions are 
fulfilled, up to utterly negligible corrections that decrease exponentially with $N$, by the 
{\it finite} version of the Coleman-Hepp model.
\vfill\eject
\centerline {\bf 1. Introductory Discussion.} 
\vskip 0.3cm
In Von Neumann\rq s [1] phenomenological picture of the measurement process, the 
coupling of a microsystem, $S$, to a measuring instrument, ${\cal I}$,  leads to the 
following two essential effects. 
\vskip 0.2cm\noindent
(I) It converts a pure state of $S$, as given by a linear combination 
${\sum}_{r=1}^{n}c_{r}u_{r}$ of its orthonormal energy eigenstates $u_{r}$, into a 
statistical mixture of these states for which ${\vert}c_{r}{\vert}^{2}$ is the probability 
of finding this system in the state $u_{r}$. This effect is often termed the \lq reduction of 
the wave packet\rq . 
\vskip 0.2cm\noindent
(II) It sends a certain set of classical, i.e. intercommuting, macroscopic variables $M$ of  
${\cal I}$ to values, indicated by pointers,  that specify which of the states $u_{r}$ of 
$S$ is actually realised. 
\vskip 0.2cm\noindent
This leaves us with the following question, which is the basic problem of the quantum 
theory of measurement.
\vskip 0.2cm\noindent
${\cal Q}$ {\it Can the standard quantum dynamics of the conservative composite 
$S_{c}$, allied to a suitable choice of the macroscopic observables $M$, lead, in 
principle, to the effects (I) and (II)?} 
\vskip 0.2cm\noindent
This is the question that we shall address here. Evidently it is crucially pertinent to that of 
the completeness of quantum mechanics, and  while some authors (e. g. [2-4])  have 
argued in favour of an affirmative answer to it, others have taken quite different views. 
For example, Von Neumann [1] and Wigner [5] have proposed that the observation of  
the pointer of ${\cal I}$ requires another measuring instrument,  ${\cal I}_{2}$, which in 
turn requires yet another instrument, and so on, in such a way that the whole process 
involves an infinite regression ending up in the observer\rq s brain (!), while others have 
taken the view that the measurement problem cannot be resolved without modification of 
the Schroedinger dynamics of $S_{c}$,  due to {\it either} its interaction with the \lq rest 
of the Universe\rq\  [6-9] 
{\it or} a certain postulated nonlinearity that leads to a classical deterministic evolution of 
its macroscopic observables [10].
\vskip 0.2cm
As regards the main requirements of a satisfactory treatment of the question ${\cal Q}$, 
it is clear from the works of Bohr [11], Jauch [12] and Van Kampen [3] that such a theory 
demands both a characterisation of the macroscopicality of the observables $M$ and an 
amplification property of the $S-{\cal I}$ coupling whereby different microstates of $S$ 
give rise to macroscopically different states of ${\cal I}$. Evidently, this implies that the 
initial state in which ${\cal I}$ is prepared must be unstable against microscopic changes 
in the state of $S$. On the other hand, as emphasised by Whitten-Wolfe and Emch [13, 
14],  the correspondence between the microstate of $S$ and the eventual observed 
macrostate of ${\cal I}$ must be stable against macroscopically small changes in the 
initial state of this instrument, of the kind that are inevitable in experimental procedures. 
Thus, the initial state of ${\cal I}$ must be {\it metastable} by virtue of this combination 
of stability and instability properties (cf. Refs. [3, 4, 12]). 
\vskip 0.2cm   
Rigorous constructive treatments of the measurement process, which take account of the 
above considerations, have been provided by Hepp [15] and Whitten-Wolfe and Emch 
[13, 14] on the basis of models for which  ${\cal I}$ is idealised as an infinite system, i.e. 
as an infinitely extended system of particles, with finite number density. In this 
idealisation, the macroscopic observables $M$ are taken to be a set of global intensive 
ones, which necessarily intercommute. This resultant picture of ${\cal I}$ and its 
macroscopic observables corresponds to that employed for the statistical mechanical 
description of large systems in the thermodynamic limit [16-18], and it has the merit of 
sharply distinguishing between macroscopically different states, since different values of 
$M$ correspond to disjoint primary representations of the observables. Moreover, in the 
treatments of the measurement problem based on this picture, the models of Hepp [15] 
and Whitten-Wolfe and Emch [13, 14] do indeed exhibit the required reduction of the 
wave-packet and the one-to-one correspondence between the pointer position of ${\cal 
I}$  and the resultant state of $S$; and these results are stable against all localised 
perturbations of the initial state of ${\cal I}$. On the debit side, however, Hepp\rq s 
model requires an infinite time for the measurement to be effected (cf. Bell [19]), while 
although that of Whitten-Wolfe and Emch achieves its measurements in finite times, it 
does so only by dint of a physically unnatural, globally extended $S-{\cal I}$ interaction.
\vskip 0.2cm
These observations motivate us to explore the mathematical structure of the measuring 
process on the basis of the model for which ${\cal I}$ is a large but finite $N$-particle 
system, with the aim of obtaining conditions under which it yields the essential results 
obtained for the infinite model instrument, but with a finite realistic observational time.  
In fact, we have achieved this aim in a recent article [20],  in which we showed that the 
quantum mechanics of finite conservative systems does indeed contain the structures 
required for an affirmative answer to the question ${\cal Q}$, as illustrated by an explicit 
treatment of the finite version of the Coleman-Hepp model [15]. This result provides 
rigorous mathematical substantiation of the arguments of Refs. [2-4] which led to the 
same main conclusion.  A key feature of our treatment [20] was the representation  of the 
macro-observables $M$ and the pointer positions of  ${\cal I}$ within the framework 
proposed by Van Kampen [21] and Emch [22], whereby $M$ comprises a set of coarse-
grained intercommuting extensive observables and  the pointer positions of ${\cal I}$ 
correspond to their simultaneous (mutually orthogonal) eigenspaces, which play the role 
of classical phase cells. Most importantly, these cells are of astronomically large 
dimensionalities, which increase exponentially with $N$, and in the ensuing theory they 
play the role analogous to that of the irreducible representation spaces of the infinite 
systems of Refs. [13-15]. As a result, the finite system model yields the essential positive 
results of the infinite one, and moreover leads to a realisation of the measurement process 
within a finite, realistic time. To be precise, this  model exhibits the properties (I) and (II) 
up to corrections that decrease exponentially with $N$ and that are therefore utterly 
negligible by any standards of experimental physics. 
\vskip 0.2cm
The object of the present note is to describe the essential features of our treatment that 
has led to these results. We start in Section 2 by constructing the generic model of 
$S_{c}$ and formulating both the time-dependent expectation values of the observables 
of $S$ and their conditional expectation values, given the values of the 
macro-observables $M$ of ${\cal I}$, subject to the assumption that $S$ and ${\cal I}$ 
are independently prepared and then coupled together at time $t=0$. In Section 3, we 
formulate the conditions on the dynamics of the model and the structure of the macro-
observables $M$ under which it exhibits the essential properties (I) and (II) of a 
measurement process. In Section 4, we show that these conditions are fulfilled the finite 
version of the Coleman-Hepp model [15]. We conclude, in Section 5, with a brief resume 
of the picture presented here.
\vskip 0.5cm
\centerline {\bf 2. The Generic Model.} 
\vskip 0.3cm
We assume that the algebras of bounded observables of the microsystem $S$, the 
instrument  ${\cal I}$ and their composite $S_{c}=(S+{\cal I})$ are those of the 
bounded operators in separable Hilbert spaces  ${\cal H}, {\cal K}$ and ${\cal 
H}{\otimes}{\cal K}$, respectively. Correspondingly, the states of  these systems are 
represented by the density matrices in the respective spaces. The density matrices for the 
pure states are then the projection operators $P(f)$ of their normalised vectors $f$. For 
simplicity we assume that ${\cal H}$ is of finite dimensionality $n$.
\vskip 0.2cm
We assume that the macroscopic description of ${\cal I}$ pertinent to the measuring 
process is based on a chosen abelian subalgebra ${\cal M}$ of ${\cal B}$, which is 
generated by coarse-grained macroscopic observables (cf. [21, 22]): these are typically 
extensive variables of parts or the whole of ${\cal I}$. The choice of ${\cal M}$ yields a 
partition of ${\cal K}$ into the simultaneous eigenspaces ${\cal K}_{\alpha}$ of its 
elements. The subspaces ${\cal K}_{\alpha}$ of ${\cal K}$ correspond to classical 
\lq phase cells\rq , which we take to represent the macrostates of ${\cal I}$ and to be 
unequivocally indicated by the \lq pointer positions\rq\ of this instrument. Most 
importantly, the dimensionality of each of these cells is astronomically large, since it is 
given essentially by the exponential of the entropy function of the macro-observables and 
thus grows exponentially with $N$. The largeness of the phase cells is closely connected 
to the robustness of the macroscopic measurement.
\vskip 0.2cm
Since ${\cal I}$ is designed so that the pointer readings are in one-to-one correspondence 
with the eigenstates $u_{1},. \ .,u_{n}$ of $S$, we assume that the index ${\alpha}$ of 
its macrostates also runs from $1$ to $n$. Hence, denoting the projection operator for 
${\cal K}_{\alpha}$ by ${\Pi}_{\alpha}$, it follows from the above specifications that 
$${\Pi}_{\alpha}{\Pi}_{\beta}={\Pi}_{\alpha}{\delta}_{{\alpha}{\beta}},\eqno(2.1)$$ 
$${\sum}_{{\alpha}=1}^{n}{\Pi}_{\alpha}=I_{\cal K}\eqno(2.2)$$
and that each element $M$ of ${\cal M}$ takes the form 
$$M={\sum}_{{\alpha}=1}^{n}M_{\alpha}{\Pi}_{\alpha},\eqno(2.3)$$
where the $M_{\alpha}$\rq s are scalars. 
\vskip 0.2cm
We assume that $S_{c}$ is a conservative system, whose Hamiltonian operator $H_{c}$, 
in ${\cal H}{\otimes}{\cal K}$, takes the form
$$H_{c}=H{\otimes}I_{\cal K}+I_{\cal H}{\otimes}K+V,\eqno(2.4)$$
where $H$ and $K$ are the Hamiltonians of $S$ and ${\cal I}$, respectively, and  $V$ is 
the $S-{\cal I}$ interaction. Further, we assume that ${\cal I}$ is an instrument of the 
first kind [12], in that the interaction $V$ induces no transitions between the eigenstates 
of $H$. Thus, since the latter comprise an orthogonal basis set $(u_{1},. \ .,u_{n})$ of 
${\cal H}$, with energy levels $({\epsilon}_{1},. \ .,{\epsilon}_{n})$, respectively, the 
operators $H$ and $V$ take the forms ${\sum}_{r=1}^{n}{\epsilon}_{r}P(u_{r})$ and 
${\sum}_{r=1}^{n}P(u_{r}){\otimes}V_{r}$, respectively, where the $V_{r}$\rq s are 
self-adjoint operators in ${\cal K}$. Hence, by Eq. (2.4),  $H_{c}$ reduces to the form 
$$H_{c}={\sum}_{r=1}^{n}P(u_{r}){\otimes}K_{r},\eqno(2.5)$$
where
$$K_{r}=K+V_{r}+{\epsilon}_{r}I_{\cal K}.\eqno(2.6)$$
Hence the one-parameter unitary group 
${\lbrace}U_{c}(t)={\rm exp}(iH_{c}t){\vert}t{\in}{\bf R}{\rbrace}$, which governs 
the dynamics of $S_{c}$, is given by the formula 
$$U_{c}(t)={\sum}_{r=1}^{n}P(u_{r}){\otimes}U_{r}(t),
\eqno(2.7)$$
where
$$U_{r}(t)={\rm exp}(iK_{r}t).\eqno(2.8)$$
\vskip 0.2cm
We assume that the the systems $S$ and ${\cal I}$ are prepared, independently of one 
another, in initial states represented by density matrices ${\omega}$ and ${\Omega}$, 
respectively, and then coupled together at time $t=0$. Thus the initial state of the 
composite $S_{c}$ is given by the density matrix ${\omega}{\otimes}{\Omega}$ in 
${\cal H}_{c}:={\cal H}{\otimes}{\cal K}$. Further, we assume that the initial state of 
$S$ is pure, and thus that ${\omega}$ is the projection operator $P({\psi})$ for a vector 
${\psi}$ in ${\cal H}$. The initial state of $S_{c}$ is then
$${\Phi}=P({\psi}){\otimes}{\Omega}.\eqno(2.9)$$
Further, ${\psi}$ is a linear combination of the basis vectors $(u_{1},. \  .,u_{n})$ and 
hence takes the form 
$${\psi}={\sum}_{r=1}^{n}c_{r}u_{r},\eqno(2.10)$$
where
$${\sum}_{r=1}^{n}{\vert}c_{r}{\vert}^{2}=1.\eqno(2.11)$$
\vskip 0.2cm
Since the evolute at time $t \ ({\geq}0)$ of the initial state ${\Phi}$ of $S_{c}$ is 
$U_{c}^{\star}(t){\Phi}U_{c}(t):={\Phi}(t)$,  it follows from Eqs.  (2.7), (2.9) and 
(2.10) that
$${\Phi}(t)={\sum}_{r,s=1}^{n}{\overline c}_{r}c_{s}P_{r,s}
{\otimes}{\Omega}_{r,s}(t),\eqno(2.12)$$
where $P_{r,s}$ is the operator in ${\cal H}$ defined by the equation
$$P_{r,s}f=(u_{s},f)u_{r} \ {\forall} \ f{\in}{\cal H}\eqno(2.13)$$
and  
$${\Omega}_{r,s}(t)=U_{r}^{\star}(t){\Omega}U_{s}(t).\eqno(2.14)$$
\vskip 0.3cm 
{\it Expectation and Conditional Expectation Values of Observables.} The observables of 
$S_{c}$ with which we shall be concerned are just the self-adjoint elements of ${\cal 
A}{\otimes}{\cal M}$. Their expectation values for the time-dependent state ${\Phi}(t)$ 
are given by the formula
$$E\bigl(A{\otimes}M\bigr)={\rm Tr}\bigl({\Phi}(t)[A{\otimes}M]\bigr) \ {\forall} \ 
A{\in}{\cal A}, \ M{\in}{\cal M}.\eqno(2.15)$$
In particular, the expectation values of the observables of $S$ are given by the equation
$$E(A)=E(A{\otimes}I_{\cal K})  \ {\forall} A{\in}{\cal A},\eqno(2.16)$$
while the probability that the macrostate of ${\cal I}$ corresponds to the cell ${\cal 
K}_{\alpha}$ is 
$$w_{\alpha}=E(I_{\cal H}{\otimes}{\Pi}_{\alpha}).\eqno(2.17)$$
Further, in view of the abelian property of the algebra ${\cal M}$, the functional $E$ 
induces a conditional expectation $E(.{\vert}{\cal K}_{\alpha})$ on ${\cal A}$, given 
the macrostate ${\cal K}_{\alpha}$, according to the following prescription. Since, by 
Eq. (2.3), $M_{\alpha}$ is the (sharply defined) value of the observable $M$ for this 
macrostate, any such conditional expectation functional, compatible with $E$, is a 
positive normalised linear functional on ${\cal A}$ that satisfies the condition 
$$E(A{\otimes}M)={\sum}_{{\alpha}=1}^{n}E(A{\vert}{\cal K}_{\alpha}) 
M_{\alpha}w_{\alpha},      \ {\forall} \ A{\in}{\cal A}, \ M{\in}{\cal M}.\eqno(2.18)$$
Consequently, since Eqs. (2.3) and (2.15) imply that 
$$E(A{\otimes}M)={\sum}_{{\alpha}=1}^{n}E(A{\otimes}{\Pi}_{\alpha}) 
M_{\alpha},      \ {\forall} \ A{\in}{\cal A}, \ M{\in}{\cal M},$$
the only admissible form for $E(.{\vert}{\cal K}_{\alpha})$ is given by the formula 
$$E(A{\vert}{\cal K}_{\alpha})=E(A{\otimes}{\Pi}_{\alpha})/w_{\alpha} \ 
{\forall} \ A{\in}{\cal A}, \ w_{\alpha}{\neq}0.\eqno(2.19)$$
It follows immediately from this equation that the positivity, linearity and normalisation 
properties of $E(.{\vert}{\cal K}_{\alpha})$ ensue from those of $E$. 
\vskip 0.5cm
\centerline {\bf 3 The Measurement Process}
\vskip 0.3cm
Suppose now that a reading of the pointer of ${\cal I}$ is made at time $t$. Then, 
according to the standard probabilistic interpretation of quantum mechanics, it follows 
from the above specifications that
\vskip 0.2cm\noindent
(i) $E(A)$ is the expectation value of  the observable $A$ of $S$ immediately before the 
reading;
\vskip 0.2cm\noindent
(ii) $w_{\alpha}$ is the probability that the reading yields the result that the macrostate 
of ${\cal I}$ corresponds to the cell ${\cal K}_{\alpha}$; and,
\vskip 0.2cm\noindent
(iii) in that case,  $E(A{\vert}{\cal K}_{\alpha})$ is the expectation value of $A$ 
immediately after the measurement.
\vskip 0.2cm\noindent
Thus, the measurement process is governed by the forms of the functionals $E$ and 
$E(.{\vert}{\cal K}_{\alpha})$ on the algebra ${\cal A}.$  Further, defining
$$F_{r,s:{\alpha}}=Tr\bigl({\Omega}_{r,s}(t){\Pi}_{\alpha}\bigr) \ 
{\forall} \ r,s,{\alpha}{\in}{\lbrace}1,. \ .,n{\rbrace},\eqno(3.1)$$
it follows from Eqs. (2.2), (2.12)-(2,14), (2.16), (2.19) and (3.1) that these functionals are 
determined by the form of $F$ according to the equations
$$E(A)={\sum}_{r=1}^{n}{\vert}c_{r}{\vert}^{2}(u_{r},Au_{r})+
{\sum}_{r{\neq}s;r,s=1}^{n}{\sum}_{{\alpha}=1}^{n}F_{r,s:{\alpha}}
{\overline c}_{r}c_{s}(u_{r},Au_{s}) \ {\forall} \ A{\in}{\cal A}\eqno(3.2)$$
and
$$E(A{\vert}{\cal K}_{\alpha})={\sum}_{r,s=1}^{n} F_{r,s;{\alpha}}
{\overline c}_{r}c_{s}(u_{r},Au_{s}) /w_{\alpha}\ {\forall} \ A{\in}
{\cal A}, \ w_{\alpha}{\neq}0.\eqno(3.3)$$
Key properties of $F$ that follow from Eqns. (2.2), (2.14) and (3.1) are that
$${\sum}_{{\alpha}=1}^{n}F_{r,r:{\alpha}}=1,\eqno(3.4)$$
$$1{\geq}F_{r,r:{\alpha}}{\geq}0,\eqno(3.5)$$
$$F_{r,s:{\alpha}}={\overline F}_{s,r:{\alpha}},\eqno(3.6)$$
and that, for $z_{1},. \ .,z_{n}{\in}{\bf C}$, the sesquilinear form ${\sum}_{r,s=1}^{n}
{\overline z}_{r}z_{s}F_{r,s;{\alpha}}$ is positive. Hence
$$F_{r,r;{\alpha}}F_{s,s;{\alpha}}{\geq}{\vert}F_{r,s;{\alpha}}{\vert}^{2}.
\eqno(3.7)$$
\vskip 0.3cm 
{\it The Ideal Instruments.} We term the instrument ${\cal I}$ {\it ideal} if  there is a 
one-to-one correspondence between the pointer reading ${\alpha}$ and the eigenstate 
$u_{r}$ of $S$, on a realistic observational time scale. Thus ${\cal I}$ is ideal if, for 
times $t$ greater than some critical value, ${\tau}$, and less, in order of magnitude, than 
the Poincare' recurrence times,
\vskip 0.2cm\noindent
(I.1) the time-dependent state ${\Omega}_{r,r}(t)$ of ${\cal I}$, that arises in 
conjunction with the state $u_{r}$ of $S$ in the formula (2.14), lies in one of the 
subspaces ${\cal K}_{\alpha}$;
\vskip 0.2cm\noindent
(I.2) the correspondence between $r$ and ${\alpha}$ here is one-to-one, i.e. 
${\alpha}=a(r)$, where $a$ is an invertible transformation $a$ of the point set 
${\lbrace}1,. \ .,n{\rbrace}$; and 
\vskip 0.2cm\noindent
(I.3) this correspondence is stable with respect to perturbations of the initial state 
${\Omega}$ of ${\cal I}$ that are localised, in the sense that each of them leaves this 
state unchanged outside some region contained in a ball of volume $O(1)$ with respect to 
$N$. 
\vskip 0.2cm\noindent
The conditions (I.1) and (I.2) signify that, for times $t$ in the range specified there,
$${\rm Tr}\bigl({\Omega}_{r,r}(t){\Pi}_{\alpha}\bigr)={\delta}_{a(r),{\alpha}},$$
i.e., by Eq. (3.1),
$$F_{r,r:{\alpha}}={\delta}_{a(r),{\alpha}}.\eqno(3.8)$$
Further, by Eqs. (3.4) and (3.5) and the invertibility of the function $a$, Eq. (3.8) is 
equivalent  to the condition
$$F_{r,r;a(r)}=1.\eqno(3.8)^{\prime},$$
while,  by Eqs. (3.7) and (3.8) and the invertibility of $a$,
$$F_{r,s;{\alpha}} =0 \ {\rm for} \ r{\neq}s.\eqno(3.9)$$
Hence, by Eqs. (3.2) and (3.9),
$$E(A)={\sum}_{r=1}^{n}{\vert}c_{r}{\vert}^{2}(u_{r},Au_{r}),\eqno(3.10)$$
which signifies that, {\it immediately before} the pointer reading, $S$ is in the mixed 
state represented by the density matrix
$${\rho}={\sum}_{r=1}^{n}{\vert}c_{r}{\vert}^{2}P(u_{r}).\eqno(3.11)$$
Thus we have a reduction of the wave packet, as given by the  transition from the pure 
state ${\psi} \ (={\sum}_{r=1}^{n}c_{r}u_{r})$ to this mixed state 
${\rho}$. 
\vskip 0.2cm
Moreover, since $E(I_{\cal H}{\vert}{\cal K}_{\alpha}){\equiv}1$, it follows from Eqs. 
(3.3), (3.8) and (3.9) that
$$w_{a(r)}{\equiv}w_{a(r)}E(I_{\cal H}{\vert}{\cal K}_{a(r)})=
{\vert}c_{r}{\vert}^{2}\eqno(3.12)$$ 
and
$$E(A{\vert}{\cal K}_{a(r)})= (u_{r},Au_{r}).\eqno(3.13 )$$
Thus, in view of the definition (2.17) of $w_{\alpha}$, Eq. (3.12) signifies that 
${\vert}c_{r}{\vert}^{2}$ is the probability that the pointer reading of ${\cal I}$ 
corresponds to the macrostate represented by the cell ${\cal K}_{a(r)}$, while Eq. (3.13) 
signifies that the state of $S$ {\it immediately after} this reading is the pure vectorial one 
$u_{r}$. Thus the property (3.8) implies both the reduction of the wave packet of $S$ 
and the one-to-one correspondence between the pointer position of ${\cal I}$ and the 
resultant state of $S$. \vskip 0.3cm 
{\it  Normal Measuring Instruments.} We term the instrument ${\cal I}$ 
{\it  normal} if  the following conditions are fulfilled.
\vskip 0.2cm\noindent
(N.1) A weaker form of the ideality condition (3.8),  or equivalently (3.8)$^{\prime}$, 
prevails, to the effect that the difference between the two sides of the latter formula is 
negligibly small, i.e., noting Eq. (3.4), that 
$$0<1-F_{r,r;a(r)}<{\eta}(N),\eqno(3.14)$$  
where, for large $N, \ {\eta}(N)$  is miniscule by comparison with unity: in the case of 
the finite version of the Coleman-Hepp model treated in Section 4, it is ${\exp}(-cN)$, 
where $c$ is a fixed positive constant of the order of unity. We note that, by Eq. (3.4) and 
the positivity of ${\Pi}_{\alpha}$, the condition (3.14) is equivalent to the inequality
$$0<{\sum}_{r{\neq}a^{-1}({\alpha})}F_{r,r;{\alpha}}<{\eta}(N).
\eqno(3.14)^{\prime}$$
Further, it follows from Eqs. (3.7), (3.14) and (3.14)$^{\prime}$ that 
$${\vert}F_{r,s;{\alpha}}{\vert}<{\eta}(N)^{1/2} \ 
{\rm for} \ r{\neq}s.\eqno(3.15)$$
\vskip 0.2cm\noindent 
(N.2) This condition (N.1) is stable under localised modifications of the initial state 
${\Omega}$ of ${\cal I}$. This stability condition may even be strengthened to include 
global perturbations of ${\Omega}$ corresponding to small changes in its intensive 
thermodynamic parameters (cf. the treatment of the Coleman-Hepp model in Section 4).
\vskip 0.2cm 
It follows now from Eqs. (3.2), (3.3) and (3.15) that the replacement of the ideal 
condition (3.8) by the normal one (3.14)  leads to modifications of the order 
${\eta}(N)^{1/2}$ to the formulae (3.10) and (3.13). In particular, it implies that a 
pointer reading ${\alpha}$ signifies that it is overwhelmingly probable, but not 
absolutely certain, that the state of $S$ is 
$u_{a^{-1}({\alpha})}$, as the following argument shows. Suppose that the initial state 
of $S$ is $u_{r}$. Then, by Eq. (2.12), the state of $S_{c}$ at time $t$ is 
$P(u_{r}){\otimes}{\Omega}_{r,r}(t)$ and consequently, by Eqs.(3.1) and 
(3.14)$^{\prime}$, there is a probability of the order of ${\eta}(N)$ that the pointer 
reading is given by a value ${\alpha}$, different from $a(r)$, of the indicator parameter 
of ${\cal I}$. In the freak case that this possibility is realised, this would mean that the 
state $u_{r}$ of $S$ led to a pointer reading ${\alpha}{\neq}a(r)$. Hence, in this case, 
any inference to the effect that a pointer reading ${\alpha}$ signified that the state of $S$ 
was $u_{a^{-1}({\alpha})}$ would be invalid. 
\vskip 0.3cm
{\it Comment.} The distinction between ideal and normal instruments is essentially 
mathematical rather than observational, since for an instrument of the latter kind, the 
odds against  the pointer indicating a  \lq wrong\rq\  microstate of $S$ are overwhelming. 
\vskip 0.5cm
\centerline {\bf 4. The Finite Coleman-Hepp Model.} 
\vskip 0.3cm 
This model is a caricature of an electron that interacts with a finite spin chain that serves 
to measure the electronic spin [15]. In order to fit this model into the scheme of the 
previous Sections, we regard the electron, ${\cal P}$, as the composite of its spin, ${\cal 
P}_{1}$, and its orbital component, ${\cal P}_{2}$. We then take the system $S$ to be 
just ${\cal P}_{1}$ and the instrument ${\cal I}$ to be the composite of ${\cal P}_{2}$ 
and the chain ${\cal C}$. Thus, we build the model of $S_{c}=(S+{\cal I})$ from its 
components in the following way.
\vskip 0.3cm
{\it The System $S={\cal P}_{1}$.} This is just a single Pauli spin. Thus, its state space 
is ${\cal H}={\bf C}^{2}$ and its three-component spin observable is given by the Pauli 
matrices $(s_{x},s_{y},s_{z})$. We denote by $u_{\pm}$ the eigenvectors of $s_{z}$ 
whose eigenvalues are ${\pm}1$, respectively. These vectors then form a basis in ${\cal 
H}$. We denote their projection operators by $P_{\pm}$, respectively.
\vskip 0.3cm
{\it The System ${\cal I}=({\cal P}_{2}+{\cal C})$.} We assume that ${\cal P}$ moves 
along, or parallel to, the axis $Ox$ and thus that the state space of ${\cal P}_{2}$ is 
${\tilde {\cal K}}:=L^{2}({\bf R})$. We assume that ${\cal C}$ is a chain of Pauli spins 
located at the sites $(1,2,. \  .,2L+1)$, of  $Ox$, where $L$ is a positive integer. Thus, the 
state space of ${\cal C}$ is ${\hat {\cal K}}:=({\bf C}^{2})^{(2L+1)}$, and therefore 
that of ${\cal I}$ is ${\cal K}={\tilde {\cal K}}{\otimes}{\hat {\cal K}}$. 
\vskip 0.2cm
The spin at the site $n$ of ${\cal C}$ is represented by Pauli matrices 
$({\sigma}_{n,x},{\sigma}_{n,y},{\sigma}_{n,z})$ that act on the $n$\rq th 
${\bf C}^{2}$ component of ${\hat {\cal K}}$. Thus, they may be canonically identified 
with operators in  ${\hat {\cal K}}$ that satisfy the standard Pauli relations
$${\sigma}_{n,x}^{2}={\sigma}_{n,y}^{2}={\sigma}_{n,z}^{2}={\hat I}; \ 
{\sigma}_{n,x}{\sigma}_{n,y}=i{\sigma}_{n,z}, \ {\rm etc},\eqno(4.1)$$
together with the condition that the spins at different sites intercommute.
\vskip 0.2cm
We assume that ${\cal P}_{1}, \ {\cal P}_{2}$ and ${\cal C}$ are independently 
prepared before being coupled together at time $t=0$. Further, we assume that the initial 
states of ${\cal P}_{1}$ and ${\cal P}_{2}$ are pure ones, represented by vectors 
${\psi}$ and  ${\phi}$ in ${\cal H}$ and ${\tilde {\cal K}}$, respectively, while that of 
${\cal C}$ is given by a density matrix ${\hat {\Omega}}$, in  ${\hat {\cal K}}$.Thus, 
the initial state of ${\cal  I}$ is
$${\Omega}=P({\phi}){\otimes}{\hat {\Omega}},\eqno(4.2)$$
where $P({\phi})$ is the projection operator for ${\phi}$. We assume that ${\phi}$ has 
support in a finite interval $[c,d]$ and that ${\hat {\Omega}}$ takes the form
$${\hat {\Omega}}={\otimes}_{n=1}^{2L+1}{\hat {\omega}}_{n},\eqno(4.3)$$
where ${\hat {\omega}}_{n}$, the initial state of the $n$\rq th spin of ${\cal C}$, is give 
by the formula
$${\hat {\omega}}_{n}={1\over 2}(I_{n}+m{\sigma}_{n,z}),\eqno(4.4)$$
where $0<m{\leq}1$. Thus, asuming that there are no interactions between the spins of 
${\cal C}, \ {\hat {\Omega}}$ is the equilibrium state obtained by subjecting this chain 
to a certain temperature-dependent magnetic field, directed along $Oz$. $m$ is then the 
magnitude of the resultant polarisation of this chain. One sees immediately from Eqs. 
(4.3) and (4.4) that ${\hat {\Omega}}$ is a pure state if $m=1$: otherwise it is mixed. 
\vskip 0.3cm
{\it The Dynamics.} Following Hepp [15], we assume that the Hamiltonian for the 
composite system $S_{c}$ is
$$H_{C}=I_{\cal H}{\otimes}p{\otimes}I_{\hat {\cal K}}+
P_{-}{\otimes}{\sum}_{n=1}^{2L+1}V(x-n){\otimes}{\sigma}_{n,x},\eqno(4.5)$$
where $p$ and $V(x-n)$ are the differential and multiplicative operators in 
$L^{2}({\bf R}) \ (={\tilde {\cal K}})$ that transform $f(x)$ to $-i{\hbar}df(x)/dx$ and 
$V(x-n)f(x)$, respectively, and $V$ is a bounded, real valued function on ${\bf R}$ with 
support in a finite interval $[a,b]$. Thus, in the notation of Eq. (2.6), but with $r$ taking 
just the values ${\pm}$,
$$K_{+}=p{\otimes}I_{\hat {\cal K}}  \ {\rm and} \ K_{-}=
p{\otimes}I_{\hat {\cal K}}+
{\sum}_{n=1}^{2L+1}V(x-n){\otimes}{\sigma}_{n,x}.\eqno(4.6)$$
The assumption here that the Hamiltonian for the free motion of ${\cal P}$ is linear 
rather than quadratic in $p$ serves to simplify the model by avoiding dispersion of the \lq 
electronic wave packet\rq\ .
\vskip 0.2cm 
The unitary groups $U_{\pm}$ generated by $iK_{\pm}$ are given by the formula
$$U_{\pm}(t) ={\exp}(iK_{\pm}t)\eqno(4.7)$$ 
and  the evolutes of ${\Omega}$ due to the actions of $U_{\pm}(t)$ are 
$${\Omega}_{\pm}(t):=U_{\pm}^{\star}(t){\Omega}U_{\pm}(t).\eqno(4.8)$$
These states are evidently the versions, for this model, of ${\Omega}_{r,r}(t)$, as 
defined by Eq. (2.14),  with the double suffix $(r,r)$ represented by $+$ or $-$. It follows 
now from Eqs. (4.2) and (4.6)-(4.8) that
$${\Omega}_{+}(t)=P({\phi}_{t}){\otimes}{\hat {\Omega}},\eqno(4.9)$$
where
$${\phi}_{t}(x)={\phi}(x+t).\eqno(4.10)$$
As for ${\Omega}_{-}$ it is convenient to formulate its evolution in interaction 
representation, in terms of the unitary operator
$$W(t):=U_{-}(t){\rm exp}(-i[p{\otimes}I_{\hat {\cal K}}]t).\eqno(4.11)$$
Thus, by Eqs. (4.2), (4.8) and (4.11),
$${\Omega}_{-}(t)= {\rm exp}(-i[p{\otimes}I_{\hat {\cal K}}]t)
\bigl(W^{\star}(t)[P({\phi}){\otimes}{\hat {\Omega}}]W(t)\bigr)
{\rm exp}(i[p{\otimes}I_{\hat {\cal K}}]t).\eqno(4.12)$$
By Eqs. (4.6), (4.7) and (4.11),  $W(t)$ satisfies the Dyson integral equation
$$W(t)=I_{\cal K}+i\int_{0}^{t}ds{\sum}_{n=1}^{2L+1}[V(x+s-n){\otimes}
{\sigma}_{n,x}]W(s),$$
whose solution is
 $$W(t)={\rm exp}\bigl(i{\sum}_{n=1}^{2L+1}[F_{n,t}(x){\otimes}
{\sigma}_{n,x}]\bigr),\eqno(4.13)$$
where
$$F_{n,t}(x)=\int_{0}^{t}dsV(x+s-n).\eqno(4.14)$$
Further, since the supports of $V$ and ${\phi}$ are $[a,b]$ and $[c,d]$, respectively, it 
follows from these last two equations that we may replace $F_{n,t}(x)$ by 
$\int_{\bf R}dxV(x)$ when employing Eq. (4.13) in the formula (4.12), provided that 
$$d{\leq}a+1 \ {\rm and} \  t{\geq}{\tau}:=2L+1-b-c.\eqno(4.15)$$
Thus, in this case, $W(t)$ may be replaced there by $I_{{\tilde {\cal K}}}{\otimes}Z$, 
where
$$Z={\rm exp}(iJ{\sum}_{n=1}^{2L+1}{\sigma}_{n,x}){\equiv}
{\otimes}_{n=1}^{2L+1}{\rm exp}(iJ{\sigma}_{n,x})\eqno(4.16)$$
and
$$J=\int_{\bf R}dxV(x).\eqno(4.17)$$
Consequently, under the conditions (4.15), Eq. (4.12) reduces to the form
$${\Omega}_{-}(t)=P({\phi}_{t}){\otimes}Z^{\star}{\hat {\Omega}}Z,$$
where ${\phi}_{t}$ is given by Eq. (4.10). On combining this equation with Eq. (4.9), we 
see that
$${\Omega}_{\pm}(t)=P({\phi}_{t}){\otimes}{\hat {\Omega}}_{\pm},\eqno(4.18),$$
where ${\hat {\Omega}}_{\pm}$ are the {\it time-independent} states given by the 
formulae
$${\hat {\Omega}}_{+}={\hat {\Omega}} \ {\rm and} \ 
{\hat {\Omega}}_{-}=Z^{\star}{\hat {\Omega}}Z.\eqno(4.19)$$
Thus, under the conditions (4.15), the chain ${\cal C}$ takes up the steady states ${\hat 
{\Omega}}_{\pm}$ corresponding to the states $u_{\pm}$ of $S$. It should be noted 
that the critical time ${\tau}$, specified in Eq. (4.15), is essentially the time required for 
the particle ${\cal P}$ to travel from end to end of the chain ${\cal C}$. It is therefore a 
reasonable macroscopic observational time. 
\vskip 0.2cm
Further, by Eqs. (4.1)-(4.4), (4.16) and (4.19), the explict forms of the states 
${\hat {\Omega}}_{\pm}$ are given by the equations
$${\hat {\Omega}}_{+}=2^{-(2L+1)}
{\otimes}_{n=1}^{2L+1}(I_{n}+m{\sigma}_{n,z})\eqno(4.20)$$ 
and 
$${\hat {\Omega}}_{-}=2^{-(2L+1)}{\otimes}_{n=1}^{2L+1}
\bigl(I_{n}+m{\sigma}_{n,z}{\rm cos}(2J)+m{\sigma}_{n,y}{\rm sin}(2J)\bigr).
\eqno(4.21)$$
\vskip 0.3cm
{\it The Macroscopic Phase Cells of  ${\cal I}$.} We take these to be the subspaces 
${\cal K}_{\pm}$ of ${\cal K}$ corresponding to positive and negative polarizations, 
respectively, of the chain ${\cal C}$ along the $Oz$-direction. To formulate these 
subspaces precisely, we first note that the eigenvalues of the total spin of ${\cal C}$ 
along that direction, namely ${\Sigma}_{z}:={\sum}_{n=1}^{(2L+1)}{\sigma}_{z}$, 
are the odd numbers between $-(2L+1)$ and $(2L+1)$. We define ${\hat {\cal K}}_{+}$ 
(resp. ${\hat {\cal K}_{-}}$) to be the subspace of  ${\hat {\cal K}}$ spanned by the 
eigenvectors of  ${\Sigma}_{z}$ with positive (resp. negative) eigenvalues. Thus, 
denoting by ${\hat {\Psi}}$ the simultaneous eigenvector of the ${\sigma}_{n,z}$\rq s 
with eigenvalues all equal to $-1$, ${\hat {\cal K}}_{\pm}$ are the  subspaces of ${\hat 
{\cal K}}$ generated by application to ${\hat {\Psi}}$ of the monomials of order greater 
than $L$ and less than $(L+1)$, respectively, in the different ${\sigma}_{n,x}$\rq s (or 
${\sigma}_{n,y}$\rq s). We denote their projection operators by ${\hat {\Pi}}_{\pm}$,  
respectively. We then define the phase cells ${\cal K}_{\pm}$ to be the subspaces 
${\tilde {\cal K}}{\otimes}{\hat {\cal K}}_{\pm}$ of ${\cal K}$, and denote their 
respective projection operators by ${\Pi}_{\pm} \ (=I_{\tilde {\cal K}}{\otimes}
{\hat {\Pi}}_{\pm})$. 
\vskip 0.2cm
Evidently, the formulation of the subspaces ${\cal K}_{\pm}$ of ${\cal K}$ here 
corresponds to that of  the previous Sections, with ${\alpha}$ taking the values $+$ and 
$-$, and fulfills the conditions of Eqs. (2.1) and (2.2). In the treatment that follows, we 
shall take the correspondence between the phase cells of  ${\cal I}$ and the eigenstates of 
$S$  to be the mapping  $r{\rightarrow}a(r)$ of  Section 3, with $a({\pm})={\pm}$. 
Thus, the phase cells ${\cal K}_{\pm}$ are the indicators for the vector states 
$u_{\pm}$, respectively. 
\vskip 0.3cm
{\it Ideality and Normality Conditions for ${\cal I}$.} It follows now the definition of 
${\Pi}_{\pm}:=I_{\tilde {\cal K}}{\otimes}{\hat {\Pi}}_{\pm}$ that, on translating the 
ideality and normality conditions (3.8)$^{\prime}$ and (3.14)$^{\prime}$, respectively, 
into the specifications for this model and using Eqs. (3.1), (3.4) and (3.5), the former 
condition reduces to the equation
$${\rm Tr}({\hat {\Omega}}_{+}{\hat {\Pi}}_{-})= {\rm Tr}({\hat {\Omega}}_{-}
{\hat {\Pi}}_{+})=0\eqno(4.22))$$
and the latter to
$$0<Max\bigl[{\rm Tr}({\hat {\Omega}}_{+}{\hat {\Pi}}_{-}), \ 
{\rm Tr}({\hat {\Omega}}_{-}{\hat {\Pi}}_{+})\bigr]<{\hat {\eta}}(L),\eqno(4.23)$$
where
$${\hat {\eta}}(L):={\eta}(2L+1).\eqno(4.24)$$
\vskip 0.3cm
{\it  Resultant Properties of ${\cal I}$.} The following proposition establish that ${\cal 
I}$ is an ideal measuring instrument for certain special values of the parameters of the 
model $S_{c}$ and is a normal one for a wide range of those parameters. Further, in the 
latter case, ${\hat {\eta}}(L)$ is exponentially small, i.e. of the order of 
${\rm exp}(-cL)$, with $c$ a positive constant of the order of unity.
\vskip 0.3cm
{\bf Proposition} [20]. {\it Assuming  the conditions of Eq. (4.15), ${\cal I}$ has the 
following properties.
\vskip 0.2cm\noindent
(a)  If $J={\pi}/2$ and $m=1$, then ${\cal I}$  is an ideal instrument, with critical time 
${\tau}$. However, although this implies that it satisfies the local stability condition (I.3), 
it is transformed to a normal instrument by small perturbations of the global polarization 
$m$. 
\vskip 0.2cm\noindent
(b) If  $J{\in}({\pi}/4,{\pi}/2]$ and $m{\in}(-1,0)$, then ${\cal I}$ is a normal 
instrument, again with critical time ${\tau}$ and with ${\hat {\eta}}(L)={\rm exp}(-
cL)$, where $c$ is a numerical constant of the order of unity: specifically 
$c=-(1/2){\rm ln}(1-m^{2}{\rm cos}^{2}(2J))$. Moreover, in this case, the instrument is 
stable both under small perturbations of the global polarisation, $m$, and under local 
modifications of its initial state.}
\vskip 0.3cm
{\bf Sketch of Proof.} Let $v_{n,{\pm}}$ denote the eigenstate of ${\sigma}_{n,z}$ 
whose eigenvalue is ${\pm}1$. Then, by definition of 
${\hat {\Pi}}_{+}$ (resp. ${\hat {\Pi}}_{-}$), the eigenstates of this projector are the 
tensor products of $n \ v_{-}$\rq s and $(2L+1-n) \ v_{+}$\rq s (resp. $n \ v_{+}$\rq s 
and $(2L+1-n) \ v_{-}$\rq s) with $n$ running  from $0$ to $L$. Hence, by Eqs. (4.20) 
and (4.21),
$${\rm Tr}({\Omega}_{+}{\hat {\Pi}}_{-})=2^{-(2L+1)}{\sum}_{n=0}^{L} 
(1+m)^{n}(1-m)^{2L+1-n}(2L+1)!/n!(2L+1-n)!\eqno(4.25)$$
and
$${\rm Tr}({\hat {\Omega}}_{-}{\hat {\Pi}}_{+})=$$
$$2^{-(2L+1)}{\sum}_{n=0}^{L}
(1-(m){\rm cos}(2J))^{n}(1+(m){\rm cos}(2J))^{2L+1-n}
(2L+1)!/n!(2L+1-n)!\eqno(4.26)$$
It follows immediately from these equations that, in the case where  $m=1$ and $J=
{\pi}/2$, the r.h.s.\rq s of these last two equations vanish and thus the ideality condition 
(4.22) is satisfied. On the other hand, if $J{\in}({\pi}/4,{\pi}/2]$ and $0<m<1$, the 
summands on the r.h.s\rq s  of Eqs. (4.25) and (4.26) are positive for all $n{\in}[0,L]$, 
and they take their largest values at $n=L$. A simple application of Sterling\rq s formula 
to $L$ times the maximum values of these summands then yields the result that the 
normality condition (4.23) is fulfilled, with ${\hat {\eta}}(L)={\rm exp}(-cN)$, where 
$c=-(1/2){\rm ln}(1-m^{2}{\rm cos}^{2}(2J))$.
\vskip 0.2cm
Further, an extension of this analysis establishes that the conditions for both ideal and 
normal behaviour of ${\cal I}$ are stable under strictly localised perturbations of the 
initial state of the instrument. On the other hand, it is immediately evident that the 
normality conditions $m{\in}(0,1), \ J{\in}({\pi}/4,{\pi}/2]$ are stable under small 
perturbations of the global polarisation $m$ that leave this quantity in the range $(0,1]$, 
whereas such perturbations change the ideality conditions $m=1, \ J={\pi}/2$ to those of 
normality. 
\vskip 0.5cm 
\centerline {\bf  5. Concluding Remarks}
\vskip 0.3cm
The realisation of  the general scheme of Sections 2 and 3 by the model of Section 4. 
signifies that the traditional quantum mechanics of finite conservative systems provides a 
perfectly adequate framework for the quantum theory of measurement. This  theory 
therefore requires no extraneous elements, such as the interaction of  $S_{c}$ with the \lq 
rest of the Universe\rq\  or a nonlinear modification of  its Schroedinger dynamics, as has 
been proposed by some authors [6-10]. Furthermore, the treatment of the model of 
Section 4 provides a clear illustration of the mathematical dichotomy of ideal and normal 
measuring instruments. It also establishes that, from an empirical standpoint, there is 
effectively no distinction between these two classes of instruments, since the odds against 
the indication by a normal instrument, of a \lq wrong\rq\ state of  a the microsystem are 
truly astronomical, being of the order of ${\rm exp}(cL)$ to one, where $c$ is of the 
order of unity. 
\vskip 0.5cm
\centerline {\bf References.} 
\vskip 0.3cm\noindent
[1] J. Von Neumann: {\it Mathematical Foundations of Quantum Mechanics}, Princeton 
University Press, Princeton, NJ, 195
\vskip 0.2cm\noindent
[2] A. Peres: Am. J. Phys. {\bf 54}, 688 (1986).
\vskip 0.2cm\noindent 
[3] N. G. Van Kampen: {\it Physica} A {\bf 153}, 97 (1988).
\vskip 0.2cm\noindent
[4] A. E. Allahverdyan, R. Balian and Th. M. Nieuwenhuizen. Eur. Phys. Lett. {\bf 61}, 
452 (2003)
\vskip 0.2cm\noindent
[5] E. P.Wigner: Pp. 171-84 of {\it Symmetries and Reflections}, Indiana University 
Press, Bloomington, 1967.
\vskip 0.2cm\noindent
[6]  N. Gisin: {\it Phys. Rev. Lett.} {\bf 52}, 1657 (1984).
\vskip 0.2cm\noindent
[7] E. Joos and H. D. Zeh: {\it Z. Phys.} B {\bf 59}, 223 (1985).
\vskip 0.2cm\noindent
[8] L. Diosi: {\it J. Phys.} A {\bf 21}, 2885 (1988).
\vskip 0.2cm\noindent
[9] I. Percival: {\it Quantum State Diffusion}, Cambridge Univ. Press,  Cambridge, 1998.
\vskip 0.2cm\noindent
[10] G. C. Ghirardi, A. Rimini and T. Weber: {\it Phys. Rev.} D {\bf 34}, 470 (1986).
\vskip 0.2cm\noindent
[11]  N. Bohr: {\it Discussion with Einstein on epistomological problems in atomic 
physics}, Pp. 200-241 of  {\it Albert Einstein: Philosopher-Scientist}, Ed. P. A. Schilp, 
The Library of Living Philosophers, Evanston, IL, 1949.
\vskip 0.2cm\noindent
[12] J, M. Jauch: {\it Foundations of Quantum Mechanics}, Addison Wesley, Reading, 
MA, 1968.
\vskip 0.2cm\noindent
[13] B. Whitten-Wolfe and G. G. Emch: {\it Helv. Phys. Acta} {\bf 49}, 45 (1976).
\vskip 0.2cm\noindent
[14] G. G. Emch: Pp. 255-264 of  {\it Quantum Information and Communication},  E. 
Donkor, A. R. Pirich and H. E. Brandt, Eds., Intern. Soc. Opt. Eng. (SPIE) Proceedings 
5105 (2003).
\vskip 0.2cm\noindent
[15] K.Hepp: {\it Helv. Phys. Acta} {\bf 45}, 237 (1972).
\vskip 0.2cm\noindent
[16] D. Ruelle: {\it Statistical Mechanics}, W. A. Benjamin, New York, 1969.
\vskip 0.2cm\noindent
[17] G. G. Emch: {\it Algebraic methods in Statistical Mechanics and Quantum Field 
Theory},  Wiley, New York, 1972.
\vskip 0.2cm\noindent
[18] G. L. Sewell: {\it Quantum Mechanics and its Emergent Macrophysics}, Princeton 
University Press, Princeton, 2002.
\vskip 0.2cm\noindent
[19]  J. S. Bell: {\it Helv. Phys. Acta} {\bf 48}, 93 (1975).
\vskip 0.2cm\noindent
[20] G. L. Sewell: {\it Rep. Math. Phys.} {\bf 56}, 271, 2005. 
\vskip 0.2cm\noindent
[21] N. G. Van Kampen: {\it Physica} {\bf 20}, 603 (1954).
\vskip 0.2cm\noindent
[22] G. G. Emch: {\it Helv. Phys. Acta} {\bf 37}, 532 (1964).
\end